\documentclass[amstex]{article}
\usepackage{graphicx}
\usepackage{dcolumn}
\usepackage{amsmath}
\usepackage{amssymb}
\usepackage{amsfonts}
\usepackage{amscd}

\newtheorem{theorem}{Theorem}
\newtheorem{lmm}[theorem]{Lemma}
\newtheorem{cor}[theorem]{Corollary}
\newtheorem{pro}[theorem]{Proposition}
\newtheorem{df}[theorem]{Definition}
\newtheorem{rmk}[theorem]{Remark}
\newtheorem{ass}[theorem]{Assumption}

\newtheorem{example}[theorem]{Example}

\newcommand\calA{{\cal A}}

\newcommand\calD{{\cal D}}
\newcommand\calE{{\cal E}}

\newcommand\calV{{\cal V}}
\newcommand\calW{{\cal W}}

\newcommand\bfZ{\bf Z}

\newcommand{\abs}[1]{\left|#1\right|}
\newcommand{\norm}[1]{\left\Vert#1\right\Vert}

\begin{document}
\newpage\thispagestyle{empty}
{\topskip 2cm
\begin{center}
{\Huge\bf
BEC of Free Bosons 
\\
on Networks}
\\
\bigskip\bigskip
\bigskip\bigskip
\bigskip\bigskip
{\Large Taku Matsui}
\\
\bigskip
{\it Graduate School of Mathematics
\\
Kyushu University
\\
1-10-6 Hakozaki, Fukuoka 812-8581
\\
JAPAN
\\
 e-mail: matsui@math.kyushu-u.ac.jp}
\\
\bigskip\bigskip\textsc{}
September ,2005
\end{center}
\bigskip\bigskip\bigskip\bigskip
\bigskip\bigskip\bigskip\bigskip
\bigskip\bigskip\bigskip\bigskip
{\bf Abstract:}  We consider free Bosons hopping on a network(infinite graph). The condition for 
Bose-Einstein condensation is given in terms of the random walk on a graph. 
In case of periodic lattices, we also consider Boson moving in an external periodic potential
and  obatined the criterion for Bose-Einstein condensation.

\newpage
\section{Results}
\setcounter{theorem}{0}
\setcounter{equation}{0}
In this note, we consider the free Bosons hopping on  networks
(general graphs) and we consider the condition for Bose-Einstein condensation.
Let $\Gamma =\{ \calV, \calE\}$ be an infinite graph where
$\calV$ is the set of vertices and $\calE$ is the set of edges.
Here for simplicity, we assume that a pair of vertices $i$ and $j$ is not connected 
by  more than two edges({\it no multiple edge}), thus
we can denote the unique edge connecting vertices $i$ and $j$ by $(i,j)$. 
The graphs we consider is connected and not oriented.
Furthermore for simplicity of presentation, we assume that 
the graph does not contain any self-loop, i.e. $(i,i) \notin \calE$.

Next we introduce free Bosons hopping on vertices of  $\Gamma$.
Let $a_j$ and $a_k^*$ ( $j$, $k$ in $\calV$)
be the creation and annihilation operators
satisfying the canonical commutation relations, $[ a_j , a_k^* ]=\delta_{jk}$. 
Smeared Boson operators, $a^*(f) $ and $a(f)$,
are defined as
\begin{equation}
a^*(f) = \sum_{j \in \calV} a_k^* f_k \quad ,\quad 
 a(f) = \sum_{j \in \calV} a_k \overline{f}_k
 \label{eqn:x1}
\end{equation} 
where $f_j$ is a complex function on $\calV$.
The function $f_j$ is referred to as a test function. 
Obviously,
$$[ a(f) , a^{*}(g) ]= (f,g)_{l^{2}(\calV)} 1$$ 
where $(f,g)$ is the inner product of $l^{2}(\calV)$, 
$$(f,g)_{l^{2}(\calV)} = \sum_{j \in \calV } \overline{f}_{j} g_{j} .$$

To be definite, we choose the set of rapidly decreasing test functions on 
$\calV$ as the test function space. Rapidly decreasing functions on $\calV$  can be introduced 
as follows. By definition, the graph distance $dist(j,k)$ of two vertices  $j$, $k$ is the minimum of 
the number of edges  giving rise to a path connecting $j$, $k$.
Fix the origin $o$ of the graph and a complex function $f$ on $\calV$ is called
rapidly decreasing if
$$\sum_{j \in \calV}  dist(j,o)^n \abs{f_j}  < \infty $$
for any positive integer $n$.
The set of all rapidly decreasing functions is denoted by 
$\emph{S}$.
Set
$$W(f) = \exp(\frac{i}{\sqrt{2}}(a^*(f)+a(f))) .$$
Then,
$$W(f)W(g) = \exp (-i\frac{Im((f ,g)_{l^{2}(\calV)}) }{2} ) W(f+g) ,$$
where $Im((f ,g)_{l^{2})(\calV})$ is the imaginary part of the inner product of
$l^{2}(\calV)$.
We denote $\mathfrak A (\Gamma)$ by the Weyl CCR  $C^*$-algebra generated by $W(f)$
($ f \in \emph{S}$ ).
\par
Now we consider the free Hamiltonians on our network. On Euclidian
spaces, the time evolution of the free Boson is determined by the Laplcian
acting on the test function space. 
Two natural candidates of the graph analogue of the Laplacian are
often used. One is the discrete Laplacian and  the other is the adjacency matrix.
By  {\it degree} of  a vertex $j$ we mean the number of edges connected to
$j$  and we denote the degree of $j$ by $d(j)$.
(The degree is sometimes called {\it coordination number}.)
The discrete Laplacian $- \bigtriangleup_{\Gamma}$ on $\Gamma$ is defined by
\begin{equation}
- \bigtriangleup_{\Gamma} f (j) = d(j) f(j) - \sum_{k :  (k,j) \in \calE}  f(k) , 
\label{eqn:x2}
\end{equation} 
where the sum is taken for all vertices $k$ connected to $j$ by an edge of $\Gamma$. 
The discrete Laplacian is a positive operator on $l^{2}(\calV )$. 
The adjacency matrix  $A_{\Gamma}$ is the off-diagonal part of the discrete Laplacian,
\begin{equation}
A_{\Gamma} f (j) = \sum_{k :  (k,j) \in \calE}  f(k) .
\label{eqn:x3}
\end{equation}
Thus when the degree of the graph is a constant $d(o)$, $ - A_{\Gamma} + d(o) = - \bigtriangleup$
and there is no physical difference of these two operators. When the degree is not constant,
we  regard  the adjacency matrix  $- A_{\Gamma}$ as a Schr$\ddot{o}$dinger operator
on a graph and the degree plays a role of potential.

We introduce assumptions for graphs to describe our main result.  
\begin{ass}
(i) We suppose that the Graph  $\Gamma$ is connected
and that $\Gamma$ is the limit of an increasing sequence of finite connected subgraphs 
$\Gamma_n =\{\calV_{n} , \calE_{n} \}$, 
$$ \Gamma  = \cup_{n=1}^{\infty}  \Gamma_n = \lim_{n\to \infty} \Gamma_n .$$
(ii) We assume the F{\o}lner condition is valid:
\begin{equation}
\lim_{n\to\infty}  \frac{\abs{\partial \Gamma_{n}}}{\abs{\Gamma_{n}}} = 0 .
\label{eqn:x4}
\end{equation}
where $\abs{\Gamma_{n}}$ is the number of vertices of $\Gamma_n$
and $\abs{\partial \Gamma_{n}}$ is the size of the boundary of $\Gamma_n$.
More precisely,  $\abs{\partial \Gamma_{n}}$ is the number of vertices of
 $\Gamma_n$ connected to  the complement $\Gamma_n^{c}$ of $\Gamma_n$
 by an edge of  $\Gamma$. 
\\
(iii) The degree of the graph is bounded,  $\sup d(j) < \infty$.
\end{ass}

The conditions (i) and (ii) above are referred to as van Hove condition in statistical mechanics
and they are used to handle thermodynamical quantities of infinite volume systems. 
The condition (iii) ensures  boundedness of the discrete Laplacians and adjacency matrices.
We note that the condition (ii) excludes certain {\it non-amenable} graphs such as Cayley trees
while the condition (iii) excludes the complete graph (all pairs of vertices connected by edges)
corresponding to the mean field free model. The case of Cayley trees is studied in \cite{Cayley}.

 For a finite sub-graph $\Gamma_{n}$, the discrete Laplacian and the adjacency matrix 
are denoted by $- \bigtriangleup_{\Gamma_{n}}$ and $A_{\Gamma_{n}}$.
If the graph is finite, the constant function $1$ is the unique ground state 
for the discrete Laplacian, 
$$- \bigtriangleup_{\Gamma}1 = 0 .$$
The unicity (up to multiplicative constant )
of the ground state follows from the Perron-Frobenius theorem for positive matrices.

The quantum Hamiltonian is the second quantization $d\Gamma$ of these operators,  thus
$H = d\Gamma (- \bigtriangleup_{\Gamma}) $ or $H = d\Gamma (- A_{\Gamma}) $.
In what follows, we consider mainly  the second quantization
of the discrete Laplacian  and we set  
\begin{equation}
H_{\Gamma} =  d\Gamma (- \bigtriangleup_{\Gamma}) 
=  \sum_{j \in \calV} \{ d(j) a_j^*a_j  -  \frac{1}{2}\sum_{k: (k,j) \in \calE} ( a_j^*a_k +a_k^*a_j) \}.  
\label{eqn:x5}
\end{equation}
Then,
$$ [ H_{\Gamma} , a^*(f) ] = a^*(- \bigtriangleup_{\Gamma}f) $$
The time evolution $\alpha^{\Gamma}_t$ of observables  in the Weyl CCR $C^*$-algebra 
$\calA (\Gamma )$ on $\Gamma$ is determined by
$$\alpha^{\Gamma}_t (W(f))= W(e^{-it  \bigtriangleup_{\Gamma}} f ) .$$
For each finite sub-graph $\Gamma_n$, let  $\varphi_{\Gamma_n}^{(\beta ,\rho)}$ 
be the equilibrium state(grand canonical ensemble)
 at the inverse temperature $\beta$ with the mean density $\rho$:
\begin{eqnarray}
& & \varphi_{\Gamma_n}^{(\beta ,\rho)} (W(f))
= \exp( - \frac{1}{4} (f , \frac{1+ z_n e^{ \beta  \bigtriangleup_{\Gamma}}}
{ 1- z_n e^{ \beta  \bigtriangleup_{\Gamma}}} f))
\nonumber\\
& &\varphi_{\Gamma_n}^{(\beta ,\rho)} (a^*(f)a(g))
= ( g , \frac{ z_n e^{ \beta  \bigtriangleup_{\Gamma}}}{ 1- z_n e^{ \beta  \bigtriangleup_{\Gamma}}} f)
\label{eqn:x6}
\end{eqnarray}
where $z_n$ ($0 < z_n <1$) is determined by
\begin{equation}
\rho =  \frac{1}{\vert \Gamma_n \vert} 
 \sum_{k \in \calV_n} \varphi_{\Gamma_n}^{(\beta ,\rho)} (a_k^* a_k)
= \frac{1}{\vert \Gamma_n \vert } 
\mathrm{tr} (\frac{ z_n e^{ \beta \bigtriangleup_{\Gamma_n}}}{ 1- z_n e^{ \beta \bigtriangleup_{\Gamma_n}}})
\label{eqn:x7}
\end{equation}
$z_n$ is the exponential of the chemical potential $\mu_{n}$ , $z_n = \exp (-\beta \mu_{n})$.
\\
\\
  When $\Gamma$ is a finite graph, we define
$ l^2_0 (\calV) = \{  (f_j) \in  l^2 (\calV) \quad \vert   \quad  \sum_{j \in \calV} f_j = 0 \}$.
 By  $E_0$ we denote the projection from $l^2 (\calV)$ to 
 $l^2_0 (\calV)$.  We set $\mathrm{tr} _0 (Q) = \mathrm{tr} ( E_0Q E_0)$ and
\begin{equation}
\overline{\rho}(\beta ) = 
\sup_{0< z<1} \limsup_{ n} \frac{1}{\vert \Gamma_n \vert } 
\mathrm{tr} 
( \frac{ z e^{ \beta \bigtriangleup_{\Gamma_n}}}{ 1- z e^{ \beta \bigtriangleup_{\Gamma_n}}}) .
\label{eqn:x8}
\end{equation}
Then,
\begin{equation}
\overline{\rho}(\beta) = 
\limsup_{n \to \infty} 
\frac{1}{\vert \Gamma_{n} \vert } \mathrm{tr} _0 ( \frac{  e^{\beta\bigtriangleup_{\Gamma_{n}}}}
{ 1- e^{\beta \bigtriangleup_{\Gamma_{n}}}})
\label{eqn:x81}
\end{equation}
because
$$\mathrm{tr} ( \frac{  z e^{\beta\bigtriangleup_{\Gamma_{n}}}}
{ 1- z e^{\beta \bigtriangleup_{\Gamma_{n}}}})
= \frac{z}{1-z} +
 \mathrm{tr} _0 ( \frac{  e^{\beta\bigtriangleup_{\Gamma_{n}}}}
{ 1- e^{\beta \bigtriangleup_{\Gamma_{n}}}}) .$$
\bigskip
\noindent
\par
The following can be shown in the same way of the Bose gas on the Euclidean space. 
( See \cite{BratteliRobinsonII} \cite{Lewis} and \cite{Cannon}. )

\begin{pro}
(i) Suppose that  $\overline{\rho}(\beta ) $ is infinite.
Then, for any $\rho >0$, there exists $z_{\infty } $ and a subsequence of finite graphs
$\Gamma_{n(i)}$ such that 
$\lim_i z_{n(i)} =   z_{\infty} <1 $ and
\begin{equation}
\varphi_{\Gamma}^{(\beta ,\rho)} (W(f)) \equiv
 \lim_i \varphi_{\Gamma_n(i)}^{(\beta ,\rho)} (W(f)) =
 \exp( - \frac{1}{4} (f , \frac{1+ z_{\infty} e^{\beta \bigtriangleup_{\Gamma}}}
 { 1- z_{\infty} e^{ \beta \bigtriangleup_{\Gamma}}} f)).
 \label{eqn:x9}
\end{equation}
\begin{equation}
 \varphi_{\Gamma}^{(\beta ,\rho)} (a^*(f)a(g)) =
\lim_n \varphi_{\Gamma_n}^{(\beta ,\rho)} (a^*(f)a(g))
= ( g , \frac{ z_{\infty} e^{ \beta \bigtriangleup_{\Gamma}}}{ 1- z_{\infty} 
e^{ \beta \bigtriangleup_{\Gamma}}} f) .
\label{eqn:x10}
\end{equation}
(ii) Suppose that $\overline{\rho}(\beta )$ is finite. 
\\
(iia) If $\rho \leq  \overline{\rho}(\beta)$, there exists $\lim_n z_n = z_{\infty} \leq 1$ 
and a subsequence of finite graphs $\Gamma_{n(i)}$ such that
 the equations (\ref{eqn:x9}) and (\ref{eqn:x10}) are valid.
\newline
(iib) Suppose $\rho > \overline{\rho}(\beta)$.
Take any subsequence of finite graphs $\Gamma_{n(i)}$ satisfying
 \begin{equation}
\overline{\rho}(\beta) = 
\lim_{i} \frac{1}{\vert \Gamma_{n(i)} \vert } 
\mathrm{tr} _0 ( \frac{  e^{\beta\bigtriangleup_{\Gamma_{n(i)}}}}
{ 1- e^{\beta \bigtriangleup_{\Gamma_{n(i)}}}})
\label{eqn:x121}
\end{equation}
Then,
\begin{equation}
\lim_i z_{n(i)} =   1 \:\: , \:\:      \lim_{i} \frac{z_{n(i)} }{\vert \Gamma_{z_{n(i)}} \vert (1-z_{n(i)})}
= \rho - \overline{\rho}(\beta).
\label{eqn:x12}
\end{equation}
Furthermore, suppose that any  $f =(f_j)$  with compact support is in the domain
of $( - \bigtriangleup_{\Gamma})^{-1/2}$ and that
the following uniformly boundedness  of matrix elements holds. 
\begin{equation}
\sup_{i,j \in \calV} (\delta_{i}, \frac{1}{(- \bigtriangleup_{\Gamma})}\delta_{j}) < \infty.
\label{eqn:x122}
\end{equation}
Then, any rapidly decreasing $f$ and $g$ are in the domain
of $( - \bigtriangleup_{\Gamma})^{-1/2}$ (hence in the domain of 
$(1-  e^{\beta \bigtriangleup_{\Gamma}})^{-1/2} $) and,
\begin{eqnarray}
& &\varphi_{\Gamma}^{(\beta ,\rho)} (W(f)) \equiv \lim_n \varphi_{\Gamma_n}^{(\beta ,\rho)} (W(f)) 
\nonumber\\
&=& \exp (-\frac{1}{2}( \rho - \overline{\rho}(\beta)) \vert \chi_{\Gamma}(f)\vert^2 )
 \exp( - \frac{1}{4} 
 (f , \frac{1+  e^{\beta  \bigtriangleup_{\Gamma}}}{ 1-  e^{ \beta  \bigtriangleup_{\Gamma}}} f))
\label{eqn:x13}
\end{eqnarray}
\begin{eqnarray}
& &\varphi_{\Gamma}^{(\beta ,\rho)} (a^*(f) a(g)) \equiv
\lim_n \varphi_{\Gamma_n}^{(\beta ,\rho)} (a^*(f) a(g)) 
\nonumber\\
&=& ( \rho - \overline{\rho}(\beta))  \chi_{\Gamma}(f) \overline{\chi}_{\Gamma}(g)
+ ( g , \frac{ e^{ \beta \bigtriangleup_{\Gamma}}}{ 1-  e^{ \beta \bigtriangleup_{\Gamma}}} f)
\label{eqn:x14}
\end{eqnarray}
where
\begin{equation}
\chi_{\Gamma} (f) = \sum_{j \in \calV}  \:\: f_j .
\label{eqn:x15}
\end{equation}
\label{Th:bec1}
\end{pro}
\noindent
\bigskip
\par
We say  {\it Bose-Einstein condensation} occurs for the second case, $\rho > \overline{\rho}(\beta)$,
in the above proposition.
Due to the equation (\ref{eqn:x13}) and (\ref{eqn:x14}),
 we have off-diagonal long range order and the U(1) gauge symmetry breaking 
for the state $\varphi_{\Gamma}^{(\beta ,\rho)} $ with a high mean density in the following sense:
$$\lim_{dist(i,j) \to \infty}\varphi_{\Gamma}^{(\beta ,\rho)} (a_{i}^* a_{j}) =
 ( \rho - \overline{\rho}(\beta)) \ne 
 0 =  \varphi_{\Gamma}^{(\beta ,\rho)} (a_{i}^*) \varphi_{\Gamma}^{(\beta ,\rho)} (a_{j}) .$$
$\varphi_{\Gamma}^{(\beta ,\eta ,\theta )} $ is decomposed into the following factor states.
\begin{eqnarray}
\psi_{\Gamma}^{(\beta ,\eta ,\theta)} (W(f))
&= &\exp \{ i( \eta - \overline{\rho}(\beta))^{1/2}  
(e^{i \theta} \chi_{\Gamma}(f) + e^{-i \theta} \overline{\chi}_{\Gamma}(f) )\}
\nonumber\\
& &\times\exp ( - \frac{1}{4} (f , \frac{1+  e^{\beta \bigtriangleup_{\Gamma_n}}}
{ 1-  e^{\beta \bigtriangleup_{\Gamma_n}}} f))
\label{eqn:x16}
\end{eqnarray}
In another word, $\varphi_{\Gamma}^{(\beta ,\rho)} $ is an integral of
$\varphi_{\Gamma}^{(\beta ,\eta ,\theta)} $ as functions of $\eta$ and $\theta$.
\noindent
\bigskip
\par
When the one-particle Hamiltonian is a discrete Laplacian,
we present the condition for Bose-Einstein condensation in terms
of the simple random walk on $\Gamma$.
By the simple random walk, we mean that the random walk can move
only to adjacent vertices and that the probability of the jump from
one vertex $i$ to another adjacent vertex is $1/d(i)$. 
The probability of the random walk moving  to adjacent vertices are all equal.  
\begin{df}
Let $p_{N}(j)$ be the probability of the random walk starting from $j$ and
returning to $j$ at the Nth step for the first time.
Let $q_{N}(j)$ be the probability of the random walk starting from $j$ and
returning to $j$ at the Nth step.
\newline
The simple random walk on $\Gamma$ is recurrent if $\sum_{N=2}^{\infty} p_{N}(j) =1$ 
for each vertex $j$. 
The simple random walk on $\Gamma$ is transient if $\sum_{N=2}^{\infty} q_{N}(j) $ is finite for
each vertex $j$.
\end{df}
It is well known that the random walk is not recurrent if and only if it is transient.
(c.f. \cite{Woess}) For general inhomogeneous networks, we use stronger conditions 
to describe our results on Bose-Einstein condensation.
\begin{df}
(i) The simple random walk on $\Gamma$ is uniformly transient if 
\begin{equation}
\sup_{j \in \calV } \sum_{N=2}^{\infty}  q_{N}(j) < \infty .
\label{eqn:x17}
\end{equation}
(ii) The simple random walk on $\Gamma$ is uniformly recurrent if
for any large $K$ there exists $M$ such that for any $j$
\begin{equation}
K \leq \sum_{N=2}^{M}  q_{N}(j) .
\label{eqn:x17b}
\end{equation}
\end{df}
We do not know whether uniformity conditions of (\ref{eqn:x17}) and (\ref{eqn:x17b})
are used in other contexts of probability theory.
The same conditions may be expressed in other terms. 

\begin{theorem}
We consider the free Bosons on a graph $\Gamma$ satisfying previous conditions  Assumption 1.1
 and assume that $\Gamma$ contains no self-loop and no multiple edge.
Let the one-particle Hamiltonian be the discrete Laplacian $-\bigtriangleup_{\Gamma}$.
\newline 
(i) Suppose that the simple random walk on $\Gamma$ is uniformly transient.
Then, 
 $$\overline{\rho}(\beta ) < \infty $$  
 and any rapidly decreasing $f$ is in the domain of 
 $(-\bigtriangleup )^{-1/2}$,
hence, alll the assumptions of Proposition \ref{Th:bec1} (iia) and (iib) are valid and
 the Bose-Einstein condensation occurs. 
\newline
(ii)  Suppose that the simple random walk on $\Gamma$ is uniformly recurrent.
Then, 
$$\overline{\rho}(\beta ) = \infty ,$$
and  the conclusion of Proposition \ref{Th:bec1} (i) holds.
\label{Th:main}
\end{theorem}
\begin{example}[Periodic lattice]
We consider an infinite graph $\Gamma$ on which the abelian group $\bf{Z}^{\nu}$ acts graph
automorphisms. By $\tau_{k}$ we denote this $\bf{Z}^{\nu}$ action.
We assume that 
\newline
(i)  the isotropy group is trivial , namely,
for any vertex $i$,   $\tau_{k}(i) =i$ implies $k=0$. 
\newline
and 
\newline
(ii) the $\bf{Z}^{\nu}$ action is co-finite in the sense that
the quotient graph $\Gamma_{0}=\Gamma/\bf{Z}^{\nu}$ is a finite graph. 
\\
A graph satisfying these conditions (i) and (ii) is referred to as a periodic lattice and 
$\Gamma_{0}$ will be called the fundamental domain of $\Gamma$. 
 In periodic lattices, by definition,  the simple random walk is uniformly transient
  (resp. uniformly recurrent)  if it is transient (resp. recurrent).
 It is known that the random walk is transient if and only if the dimension $\nu$ is greater
 than or equal to three, $3\leq \nu$.  See \cite{PeriodicLattice}. 
In the periodic case, we can replace $\limsup$ with $\lim$ in taking infinite volume
limit and we do not have to take subsequences of finite graphs.
\end{example}
\begin{example}[Defects]
Now we remove edges from graphs and we call removed edges {\it defects}.
We denote the set of defects by $\calD$. We assume that the density of defects vanishes
$$\lim_{n\to \infty} \frac{\vert\calD \cap \calV_{n}\vert}{\vert\calV_{n}\vert} = 0.$$ 
Then, our proof shows that if the simple random walk is uniformly transient for the initial
graph, BEC occurs for the graph with defects.
If the density of defects is positive, further consideration is required
and we do not know  condition for occurrence or absence of BEC in this case.
\end{example}
\noindent
\bigskip
\par
Next we discuss the case when  the one-particle Hamiltonian is an adjacency matrix. 
The transiency of the random walk does not determine the occurence of BEC. 
Such examples are provied in  \cite{Star} and\cite{Sodano}.
The examples discussed in  \cite{Star} and\cite{Sodano} are star graphs and comb graphs
for which the simple random walk are recurrent . (See \cite{Woess}.)
In \cite{Star},  I. Brunelli, G. Giusiano, F. P. Mancini, P. Sodano and A. Trombettoni
 considered  the star graph and the adjacency matrix has a spectral gap. 
 In such situation, it is easy to see the existence of the critical density and occurrence of BEC. 
In \cite{Sodano}, R.Burioni e D. Cassi, M. Rasetti, P. Sodano and A. Vezzani
investigated BEC on the comb graph. The vertices of a comb graph are same as $\bfZ^{2}$ but some
edges are removed from  $\bfZ^{2}$. In our terminology, the density of defects is strictly positive.  
\noindent
\bigskip
\par
When the graph is a periodic lattice and the adjacency matrix is the one-particle Hamiltonian, 
we have the same results as the integer lattice, so BEC occurs
if and only if   $3\leq \nu$. For the continuous space case on $\bf{R}^{\nu}$, this follows from
a result of W.Kirsch and B.Simon in \cite{KirschSimon} (See also \cite{Zagrebnov1}) .
W.Kirsch and B.Simon considered the Schr\"odinger operator with a periodic potential
and they proved that behavior  of the density of states
in the vicinity of the bottom of energy is same as that for free Laplacian.
The argument of W.Kirsch and B.Simon works for our periodic lattice as well.
We obtain the same result when we add a periodic potential to the one-particle Hamiltonian.
\par
We set 
\begin{equation}
\Gamma_{n} = \cup_{\abs{k_i}\leq n \:\:( i=1,2,...\nu)}  \tau_{k} (\Gamma_{0})
\label{eqn:x300}
\end{equation}
By $\Gamma_{n}^{(p)}$, we denote the graph
obtained by the periodic boundary condition on $\Gamma_{n}^{(p)}$.
The vertex set of $\Gamma_{n}^{(p)}$ is same as that of $\Gamma_{n}$ and
the edge set of $\Gamma_{n}^{(p)}$ is the union of $\Gamma_{n}$ and
additional edges in the way that  there is a natural $(\bfZ_{2n+1})^{\nu}$ action on
$\Gamma_{n}^{(p)}$ compatible with the shift of $\Gamma$. 
More precisely, if  $i \in \calV_{n}$, $j \notin \calV_{n}$, $(i,j) \in \calE$ 
and there exists $m \in \bfZ^{\nu}$ such that $\abs{m}=1$ and $\tau_{m}(i) =k$ for some $k$ 
in $\calV_{n}$, $(i,k)$ is an edge of $\Gamma_{n}^{(p)}$.
Using this periodic graph $\Gamma_{n}^{(p)}$, we can introduce the one-particle
Hamiltonian with the periodic boundary condition.
\begin{theorem}
Suppose the graph $\Gamma$ is a periodic lattice and $v(j)$ is a (real) periodic function
on the set of vertices $\calV$.  
Let  $E_{n}$ (resp. $E$) be the supremum of the spectrum of $  A_{\Gamma_n^{(p)}} - v$
(resp. $A_{\Gamma} - v$)
acting on $l^{2}(\calV_{n})$ (resp.$l^{2}(\calV)$) .
Consider
\begin{equation}
h_{n} =  E_{n} - A_{\Gamma_n^{(p)}} + v , 
\label{eqn:x301}
\end{equation}
\begin{equation}
h =   E - A_{\Gamma} + v  .   
\label{eqn:x302}
\end{equation}
Let $\varphi_{n}^{(\beta ,\rho)}$ be the equilibrium state associated 
with the second quatization of $h_{n}$ with a mean particle density $\rho$. It is determined by 
\begin{eqnarray}
& &\varphi_{n}^{(\beta ,\rho)} (W(f)) 
= \exp( - \frac{1}{4}  (f , \frac{1+ z_{n} e^{-\beta h_{n} }}{ 1-   z_{n} e^{ -\beta  h_{n}}} f)) ,
\nonumber\\
& &\frac{1}{\abs{\Gamma_{n}}} \varphi_{n}^{(\beta ,\rho)} (N_{n}) 
=\rho .
\label{eqn:x30}
\end{eqnarray}
(i) Suppose that $\nu \leq 2$. BEC does not occur.
For any $\rho$ the limits $\lim_{n}  z_{n} = z <1$ and
$$\lim_{ n \to \infty} \varphi_{n}^{(\beta ,\rho)} (W(f)) =\varphi^{(\beta ,\rho)} (W(f)) 
= \exp( - \frac{1}{4}  (f , \frac{1+ z  e^{-\beta h }}{ 1-  ze^{ -\beta  h}} f )) $$
exist. 
$\varphi^{(\beta ,\rho)}$ is a translationally invariant factor state of ${\mathfrak A}(\Gamma)$.
\\
(ii)Suppose that $\nu \geq 3$. Then the rapidly decreasing functions are in the domain of
$h^{-1/2}$ and we can define the critical density $\rho_{\beta}$
 via the trace $\mathrm{tr}_{l^{2}(\calV_{0})} $ over the fundamental domain of our periodic lattice:
\begin{equation}
\rho_{\beta} =  \frac{1}{\abs{\Gamma_{0}}} \mathrm{tr}_{l^{2}(\calV_{0})}
( \frac{e^{-\beta  h}}{ 1-  e^{ -\beta  h} }) 
=
\frac{1}{\abs{\Gamma_{0}}} \sum_{ a \in \calV_{0}}
( \delta_{a}, \frac{e^{-\beta  h}}{1-  e^{ -\beta  h}} \delta_{a}) .
\label{eqn:x31}
\end{equation}
Suppose  $\rho_{\beta} \leq \rho$. Then, $\lim_{n \to\infty} z_{n} =1$ and
\begin{eqnarray}
& &\varphi_{\Gamma}^{(\beta ,\rho)} (W(f)) \equiv \lim_n \varphi_{\Gamma_n}^{(\beta ,\rho)} (W(f)) 
\nonumber\\
& &= \exp (-\frac{1}{2}( \rho - \rho (\beta)) \vert \tilde{\chi}_{\Gamma}(f)\vert^2 )
 \exp( - \frac{1}{4} 
 (f , \frac{1+  e^{- \beta  h}}{ 1-  e^{ -\beta  h}} f))
\label{eqn:32}
\end{eqnarray}
$$\varphi_{\Gamma}^{(\beta ,\rho)} (a^*(f) a(g)) 
 = ( \rho - \rho(\beta))  \tilde{\chi}_{\Gamma}(f) \overline{\tilde{\chi}_{\Gamma}(g)}
+ ( g , \frac{ e^{ -\beta h}}{ 1-  e^{ - \beta h}} f) $$
where
\begin{equation}
\tilde{\chi}_{\Gamma} (f) = \lim_{n \to \infty} \abs{\Gamma_{n}}^{1/2} (\Omega_{n} , f )_{l^{2}(\calV_{n})} .
\label{eqn:x33}
\end{equation}
and $\Omega_{n}$ is the positive normalized ground state of $h_{n}$ uniquely determined by 
\begin{equation}
h_{n}\Omega_{n} = 0 , \quad
 (\Omega_{n} , \Omega_{n})_{l^{2}(\calV_{n})} = 1.
 \label{eqn:x34}
\end{equation}
Alternatively, $\tilde{\chi}_{\Gamma} (f)$ is described by the inner product of the positive periodic
ground state $\Omega = \{\Omega (i) \}$  for the infinite volume Hamiltonian $h$.
Let $\Omega$ belong to $l^{\infty}$
satisfying the normalization condition
\begin{equation}
\sum_{i \in \calV_{0}}\abs{\Omega (i)}^{2} =1 
\label{eqn:x35}
\end{equation}
Then,
\begin{equation}
\tilde{\chi}_{\Gamma} (f)  = \sum_{j \in\calV}  \Omega (i)  f_{i} =   (\Omega , f )_{l^{2}(\calV)} , 
\label{eqn:x36}
\end{equation}
which converges for any rapidly decreasing $f$.
\label{Th:Periodic}
\end{theorem}
\noindent
\bigskip
\noindent
\bigskip
\par
Next we explain key points in our proof of  Theorem \ref{Th:main} and  \ref{Th:Periodic}.
The first point is finiteness of  the critical density $\overline{\rho}(\beta )$.
For $x >0$ and $0 < z < 1$,
$$\frac{ z e^{ - \beta x}}{ 1 - z e^{ - \beta x}} = \sum_{m=1}^{\infty} z^{m} e^{ -m \beta x} .$$
Hence, to examine finiteness of $\overline{\rho}(\beta )$, we consider the trace of Gibbs weight,
$\frac{1}{\abs{\Gamma_{n}} } \mathrm{tr}( e^{-\beta h_{n}} )$.
Let $P_{l^{2}(\calV_{n})}$ be the projection  $l^{2}(\calV )$ from to $l^{2}(\calV_{n})$ and 
$d_{n}$ (resp. $d$) be the degree of $\Gamma_{n}$(resp. $\Gamma$). 
By Trotter-Kato product formula we will see
\begin{equation}
\mathrm{tr}_{l^{2}(\calV_{n})}
( \frac{ z e^{ - \beta h_{n}} }{ 1 - z e^{ - \beta  h_{n}}}) 
\leq  \mathrm{tr}_{l^{2}(\calV)}(P_{l^{2}(\Gamma_{n})} 
\frac{ z e^{ - \beta h_{n}}}{ 1 - z e^{ - \beta  h_{n}}} P_{l^{2}(\Gamma_{n})})  
\label{eqn:x37}
\end{equation}
Then, by use of the F{\o}lner condition (\ref{eqn:x4}), we show that  we can replace $d_{n}$ with $d$
 in the right-hand side in the above inequality. 
Then, we obtain
\begin{equation}
\overline{\rho}(\beta )\leq \sup_{i \in \calV} 
\left( \delta_{i} , \frac{ z e^{-\beta h}}{ 1 - z e^{ - \beta h}} 
\delta_{i} \right)_{l^{2}(\calV)} \leq   
C \left( \delta_{i} ,  \frac{1}{h + (1-z)} \delta_{i} \right)_{l^{2}(\calV)}
\label{eqn:x38}
\end{equation}
for a positive constant $C$.
If the diagonal of the Green function is bounded uniformly in $i$, 
we obtain finiteness of the critical density $\overline{\rho}(\beta )$.
It is east to see that boundedness of  the diagonal of Green function is equivalent to 
transiency of the simple random walk. This shows the first part of Theorem \ref{Th:main}.
Another inequality can be obtained by similar way and the divergence of the Green function
on the diagonal is implied by recurrence of the simple random walk. 
\par
When the graph is a periodic lattice, the situation is simpler as we employed the periodic boundary 
condition. It is easy to see
\begin{eqnarray}
&&\lim_{n \to \infty} \frac{1}{\abs{\Gamma_{n}}}
\mathrm{tr}_{l^{2}(\Gamma_{n})} 
( \frac{ z e^{ - \beta  h_{n}}}{ 1 - z e^{ - \beta h_{n}}}) 
\nonumber\\
= &&\frac{1}{\abs{\Gamma_{0}}} \left\{ \sum_{i \in \calV_{0}}
\left( \delta_{i} , \frac{ z e^{ - \beta h}}{ 1 - z e^{ - \beta h}}  \delta_{i} \right)_{l^{2}(\Gamma)}  \right\} .
\label{eqn:x39}
\end{eqnarray}
Thus, proof of finiteness or divergence of the critical density reduces to finiteness of the diagonal
of Green functions again. We show an inequality between the Dirichlet forms attached to our discrete
Schr\"{o}dinger operators. (c.f. Lemma \ref{lmm:c8}) . This tells us that the behavior of 
the density of states at the bottom of spectrum for our discrete
Schr\"{o}dinger operators with a periodic potential is same as that for our discrete Laplacian.
\noindent
\bigskip
\par
In Theorem \ref{Th:main} we assumed a number of technical conditions on
graphs. Conditions of as no self loop and of no multiple edge
are not essential. If the graph has self loops we have only to consider
random walks which stay on the vertex with certain probability.
To handle multiple edges, we modify the definition of the adjacency matrix and
its entry is the number of edges connecting vertices labeling entries.
On the other hand, the F{\o}lner condition (\ref{eqn:x4}) plays an crucial
r\^{o}le in our argument. Even though we are unable to prove Bose-Einstein
condensation without using  the F{\o}lner condition, we believe the same result
is valid for graphs without  our F{\o}lner condition.
One such example is the Cayley tree associated with the free group.
In \cite{Cayley} M.van den Berg, T.C.Dorlas and V.B.Priezzhev proved Bose-Einstein
condensation for the free Boson hopping on  a Cayley tree.
Their argument is based on explicit computation of the spectrum of the discrete Laplacian.
\noindent
\bigskip
\par
In Section 2 we present our proof of  Theorem \ref{Th:main} and Section 3 is devoted
to  Theorem \ref{Th:Periodic}. 

\section{Proof of Theorem \ref{Th:main}.}
\setcounter{theorem}{0}
\setcounter{equation}{0}
In this section, we prove Theorem \ref{Th:main}.
Let $\calW$ be a subset of vertices $\calV$
Let us regard $l^{2}(\calW)$ as a closed subspace of $l^{2}(\calV)$ and
we denote the orthogonal projection from $l^{2}(\calW )$ to $l^{2}(\calV)$
$P_{\calW}$.
We regard the degree $d(j)$ of $\Gamma$ as a multiplication operator on $l^{2}(\calV)$
denoted by $d$, while the degree of the subgraph $\Gamma_{n}$ is denoted by
$d_{n}$. $d_{n}$ is different from $d(j)$ on the boundary of $\Gamma_{n}$.  
We set
$$\overline{d} = \sup_{j \in \calV} d_{\Gamma}(j) < \infty .$$
By $tr_{\mathfrak H}(Q)$ we denote the trace of an operator $Q$ on a Hilbert space
 $\mathfrak H$.  For simplicity, $A_{\Gamma}$(resp. $A_{\Gamma_n}$ is denoted by $A$ (resp. $A_n$).
\begin{lmm}
\begin{eqnarray}
&& \:\: \mathrm{tr}_{l^{2}(\calV_{n})} (e^{\beta\bigtriangleup_{\Gamma_{n}}})
\nonumber\\
&&\leq
 \mathrm{tr}_{l^{2}(\calV)} 
( P_{\partial\Gamma_{n}} \beta A e^{\beta \bigtriangleup_{\Gamma_{n}}} P_{\partial\Gamma_{n}} )
+  \mathrm{tr}_{l^{2}(\calV_{n})}( \exp ( P_{\Gamma^{int}_{n}} 
\beta\bigtriangleup_{\Gamma} P_{\Gamma^{int}_{n}}))
\nonumber\\
&&\leq 
\mathrm{tr}_{l^{2}(\calV)}
( P_{\partial\Gamma_{n}}\beta A e^{\beta\bigtriangleup_{\Gamma} }P_{\partial\Gamma_{n}} )
+  \mathrm{tr}_{l^{2}(\calV)}
( e^{\beta\bigtriangleup_{\Gamma_{n}} }P_{\partial\Gamma_{n}} )
\nonumber\\
& &+  \mathrm{tr}_{l^{2}(\calV)} ( P_{\calV_{n}}  e^{\beta\bigtriangleup_{\Gamma}} P_{\calV_{n}} )
\label{eqn:y1}
\end{eqnarray}
where $\Gamma^{int}_{n} $ is the interior of $\Gamma_{n}$,
 $\Gamma^{int}_{n} = \calV_{n} - \partial\Gamma_{n}$.
\label{lmm:b1}
\end{lmm}
\textbf{Proof.}
The positive discrete Laplcian $-\bigtriangleup_{\Gamma_{n}}$ 
is composed of the diagonal term (multiplication operator)  $d_{n}$ and 
the off-diagonal part  identical to the adjacency matrix $A_{n}$ of. 
The difference of the (positive) discrete Laplcian 
$-\bigtriangleup_{\Gamma_{n}}$ and $-P_{\Gamma_{n}}\bigtriangleup_{\Gamma}P_{\Gamma_{n}}$
is  the diagonal part.
Using the Trotter-Kato product formula, we obtain 
\begin{equation}
\mathrm{tr}_{l^{2}(\calV_{n})}(e^{-\beta (-\bigtriangleup_{\Gamma_{n}})} ) =
\lim_{N \to \infty} tr_{l^{2}(\calV_{n})} 
\left( \left[(1+\frac{\beta A_{n}}{N})e^{ - \beta \frac{d_{n}}{N}}\right]^{N}\right) .
\label{eqn:y2}
\end{equation}
Set 
$$X(N)= \left[(1+\frac{\beta A_{n}}{N})e^{ - \beta \frac{d_{n}}{N}}\right] \: , \quad
\bar{X}(N)= P_{\Gamma^{int}_{n}} X(N) P_{\Gamma^{int}_{n}} .$$
Then, the right-hand side of the equation (\ref{eqn:y2}) is
\begin{eqnarray}
\label{eqn:y3}
&&\:\: \mathrm{tr}_{l^{2}(\calV_{n})}(X(N)^{N}) = 
\mathrm{tr}_{l^{2}(\calV_{n})} (X(N)^{N} P_{\partial\Gamma_{n}} ) +  
\mathrm{tr}_{l^{2}(\calV_{n})}(X(N)^{N} P_{\Gamma^{int}_{n}}  ) 
\nonumber\\
= && \mathrm{tr}_{l^{2}(\calV_{n})}(X(N)^{N} P_{\partial\Gamma_{n}} ) + 
 \mathrm{tr}_{l^{2}(\calV_{n})}( X(N)^{N-1} \bar{X}(N)) 
\nonumber\\
&&  + \mathrm{tr}_{l^{2}(\calV_{n})}( X(N)^{N-1} P_{\Gamma^{int}_{n}}  X(N)P_{\partial\Gamma_{n}})
 \nonumber\\
= && \mathrm{tr}_{l^{2}(\calV_{n})} (X(N)^{N} P_{\partial\Gamma_{n}} ) + 
 \mathrm{tr}_{l^{2}(\calV_{n})} ( X(N)^{N-1} \bar{X}(N))
 \nonumber\\
 &&+  
 \mathrm{tr}( X(N)^{N-1} P_{\Gamma_{int}} \frac{\beta A}{N} e^{-\beta d_{n}}P_{\partial\Gamma_{n}}) .
\end{eqnarray}
As all the terms in (\ref{eqn:y3}) are the trace of  non-negative matrices,
\begin{eqnarray}
\label{eqn:y4}
&&\mathrm{tr}_{l^{2}(\calV_{n})}  (X(N)^{N})
\nonumber\\ 
&\leq& 
 \mathrm{tr}_{l^{2}(\calV_{n})} (X(N)^{N} P_{\partial\Gamma_{n}} ) 
+  \mathrm{tr}_{l^{2}(\calV_{n})} ( X(N)^{N-1} \bar{X}(N)) 
\nonumber\\
&&+\frac{1}{N}\mathrm{tr}_{l^{2}(\calV_{n})}
( X(N)^{N-1} P_{\Gamma_{int}} \beta A P_{\partial\Gamma_{n}} e^{-\beta d_{n}})
\nonumber\\
&\leq&
\mathrm{tr}_{l^{2}(\calV_{n})} (X(N)^{N} P_{\partial\Gamma_{n}} ) +  
\mathrm{tr}_{l^{2}(\calV_{n})} ( X(N)^{N-2} \bar{X}(N)^{2}) 
\nonumber\\
&&+ \frac{2}{N}\mathrm{tr}_{l^{2}(\calV_{n})} 
 ( X(N)^{N-1} P_{\Gamma_{int}}  \beta A P_{\partial\Gamma_{n}} e^{-\beta d_{n}})
\nonumber\\
&\leq&
\mathrm{tr}_{l^{2}(\calV_{n})}(X(N)^{N} P_{\partial\Gamma_{n}} ) +  
\mathrm{tr}_{l^{2}(\calV_{n})}(  \bar{X}(N)^{N}) 
\nonumber\\
&&
+\mathrm{tr}_{l^{2}(\calV_{n})}( X(N)^{N-1} P_{\Gamma^{int}_{n}} \beta A P_{\partial\Gamma_{n}}).
\end{eqnarray} 
Now we take $N$ to infinity in (\ref{eqn:y4}),  and we obtain
\begin{eqnarray}
\label{eqn:y5}
& & \:\: \mathrm{tr}_{l^{2}(\calV_{n})}( e^{\beta \bigtriangleup_{\Gamma_{n}}} ) 
\nonumber\\
&\leq&
\mathrm{tr}_{l^{2}(\calV_{n})}(e^{\beta \bigtriangleup_{\Gamma_{n}}} P_{\partial\Gamma_{n}}) 
+
\mathrm{tr}_{l^{2}(\calV_{n})}(e^{\beta P_{\Gamma^{int}_{n}} 
\bigtriangleup_{\Gamma_{n}} P_{\Gamma^{int}_{n}} }) 
\nonumber\\
&&+
\mathrm{tr}_{l^{2}(\calV_{n})}
( e^{\beta \bigtriangleup_{\Gamma_{n}}} P_{\Gamma^{int}_{n}} \beta A P_{\partial\Gamma_{n}})
\nonumber\\
&\leq&
\mathrm{tr}_{l^{2}(\calV_{n})}(e^{\beta \bigtriangleup_{\Gamma_{n}})} P_{\partial\Gamma_{n}}) 
+
\mathrm{tr}_{l^{2}(\Gamma^{int}_{n})}(e^{\beta P_{\Gamma^{int}_{n}} 
\bigtriangleup_{\Gamma_{n}} P_{\Gamma^{int}_{n}} }) 
\nonumber\\
&&+
\mathrm{tr}_{l^{2}(\calV_{n})}
( e^{\beta \bigtriangleup_{\Gamma_{n}}} \beta A P_{\partial\Gamma_{n}}).
\end{eqnarray}
Next we show that for any subsets of vertices $\calW_{1}$ and $\calW_{2}$
satisfying  $\calW_{1} \subset \calW_{2} \subset \calV$,
\begin{equation}
\label{eqn:y6}
\mathrm{tr}_{l^{2}(\calW_{1})}(e^{\beta P_{\calW_{1}} 
\bigtriangleup_{\Gamma_{n}} P_{\calW_{1}}}) 
\leq 
\mathrm{tr}_{l^{2}(\calW_{2})}(e^{\beta P_{\calW_{2}} 
\bigtriangleup_{\Gamma_{n}} P_{\calW_{2}}}) 
\leq
\mathrm{tr}_{l^{2}(\calW_2)}(e^{\beta  \bigtriangleup_{\Gamma}})
\end{equation}
Again, this is due to the Trotter-Kato formula and non-negativity of matrix elements of
the adjacency matrix.
\begin{eqnarray*}
& &\:\: \mathrm{tr}_{l^{2}(\calW_{1})}(e^{\beta P_{\calW_{1}} 
\bigtriangleup_{\Gamma_{n}} P_{\calW_{1}}}) 
=\lim_{N\to \infty} 
\mathrm{tr}_{l^{2}(\calW_{1})}(( P_{\calW_{1}}(1+\frac{\beta A_{n}}{N})
P_{\calW_{1}}e^{ - \beta d_n / N} )^{N}) 
\\
&&\leq \lim_{N\to \infty} 
\mathrm{tr}_{l^{2}(\calW_{2})}(( P_{\calW_{2}}(1+\frac{\beta A_{n}}{N})
P_{\calW_{2}}e^{ - \beta d_n /N} )^{N}) 
=\mathrm{tr}_{l^{2}(\calW_{2})}(e^{\beta P_{\calW_{2}}\bigtriangleup_{\Gamma_{n}} P_{\calW_{2}}})
\nonumber\\
&&\leq \lim_{N\to \infty} 
\mathrm{tr}_{l^{2}(\calW_{2})}(((1+\frac{\beta A}{N} )e^{- \beta d_n} )^{N})
= \mathrm{tr}_{l^{2}(\calW_2)}(e^{- \beta ( d_n - A)})
\end{eqnarray*}
As $n$ is arbitrary, we take $n$ to $\infty$ and we obtain,
$$ \mathrm{tr}_{l^{2}(\calW_{2})}(e^{\beta P_{\calW_{2}} \bigtriangleup_{\Gamma_{n}} P_{\calW_{2}}}) 
\leq
\mathrm{tr}_{l^{2}(\calW_2)}(e^{- \beta ( d - A)}) = 
\mathrm{tr}_{l^{2}(\calW_2)}(e^{\beta  \bigtriangleup_{\Gamma}})$$
Combined with (\ref{eqn:y6}) , the equation (\ref{eqn:y5}) implies (\ref{eqn:y1}).
\textbf{End of Proof.}
\\
\\
\begin{lmm}
There exists a constant $C$ such that
\begin{equation}
\mathrm{tr}_{l^{2}(\calV_{n})} (\frac{ ze^{\beta\bigtriangleup_{\Gamma_{n}}}}
{1-ze^{\beta\bigtriangleup_{\Gamma_{n}}}})
\leq  C \vert \partial \Gamma_{n}\vert  ( \beta \frac{z}{(1-z)^{2}} +  \frac{z}{1-z} ) +
tr_{l^{2}(\calV)} (P_{\Gamma_{n}} 
\frac{ ze^{\beta\bigtriangleup_{\Gamma}} }{ 1-ze^{\beta\bigtriangleup_{\Gamma}}} 
P_{\Gamma_{n}})
\label{eqn:y7}
\end{equation}
\label{lmm:b2}
\end{lmm}
\textbf{Proof.}
As the heat kernels  are contractive, $e^{\beta\bigtriangleup_{\Gamma}}$ and 
$e^{\beta\bigtriangleup_{\Gamma_{n}}}$ have the norm $1$.
By definition,
$$ \mathrm{tr}_{l^{2}(\calV)}( P_{\partial\Gamma_{n}}) = \vert \partial \Gamma_{n}\vert .$$
Thus, we have
\begin{equation}
\abs{ \mathrm{tr}_{l^{2}(\calV)}( P_{\partial\Gamma_{n}}\beta A 
e^{\beta\bigtriangleup_{\calV}}P_{\partial\Gamma_{n}} )}
\leq  \beta \vert \partial \calV_{n}\vert  \beta \norm{A}, 
\label{eqn:y71}
\end{equation}
and
\begin{equation}
\mathrm{tr}_{l^{2}(\Gamma)} ( e^{\beta\bigtriangleup_{\Gamma_{n}} }P_{\partial\Gamma_{n}} )
\leq  \vert \partial \Gamma_{n}\vert .
\label{eqn:y72}
\end{equation}
On the other hand, recall that 
$$\frac{ ze^{\beta\bigtriangleup_{\Gamma_{n}}}}{1-ze^{\beta\bigtriangleup_{\Gamma_{n}}}}
= \sum_{M=1}^{\infty}  z^{M} e^{ M \beta\bigtriangleup_{\Gamma_{n}}} $$
for $0<z<1$ and $\beta >0$.
We combine the trace of this geometric series with the inequality (\ref{eqn:y1})
(\ref{eqn:y71}) and (\ref{eqn:y72}) to get (\ref{eqn:y7}). \textbf{End of Proof.}
\\
\\
\begin{lmm}
(i) Suppose that the simple random walk is transient. Then,
the delta function $\delta_{j}$ is in the domain of $-\bigtriangleup_{\Gamma}^{-1/2}$
\\
(ii) Suppose further that  the simple random walk is uniformly transient. Then,
\begin{equation}
\left(\delta_{i} , \frac{1}{-\bigtriangleup_{\Gamma}}   \delta_{j}  \right)_{l^{2}(\Gamma)} 
\leq C r^{(0)}_{ij}
\label{eqn:y8}
\end{equation}
where $ r^{(0)}_{ij}$ is the probability of the simple random walk starting from
$i$ arriving at $j$.
\label{lmm:b3}
\end{lmm}
\textbf{Proof.}
(i)
Note that
\begin{equation}
q_{N}(j)  = \left(\delta_{i} ,  (\frac{1}{d}A_{\Gamma})^{N} \delta_{i} \right)_{l^{2}(\Gamma)} 
\label{eqn:y9}
\end{equation}
Then, for positive $z$, we have the following Neuman expansion:
\begin{eqnarray}
&& \left(\delta_{i} ,  \frac{1}{z - \bigtriangleup_{\Gamma}} \delta_{i} \right)_{l^{2}(\Gamma)} 
 =\left(\delta_{i} ,  \frac{1}{d+z -A_{\Gamma}} \delta_{i} \right)_{l^{2}(\Gamma)} 
 \nonumber\\
&&=\sum_{N=0}^{\infty}
 \left(\delta_{i} ,  (\frac{1}{d+z}A_{\Gamma})^{N}  \frac{1}{d+z} \delta_{i} \right)_{l^{2}(\Gamma)}
 \nonumber\\ 
& &\leq (1 + \sum_{N=2}^{\infty} q_{N}(i) \frac{1}{d(i)})  < \infty .
\label{eqn:y10}
\end{eqnarray}
(ii) Let $r_{ij}(N)$ be the probability of the random walk starting from $i$ and arriving at $j$
at the Nth step and let $r_{ij}^{(0)}(N)$ be he probability of the random walk starting from $i$ and 
arriving at $j$ at the Nth step for the first time. Then,
\begin{equation}
r_{ij}(N) = r_{ij}^{(0)}(N) + \sum_{k= dist(i,j)}^{N-2}  r_{ij}(k)  p_{j}(N-k) .
\label{eqn:y11}
\end{equation}
Set
\begin{eqnarray*}
& &\overline{r}_{ij}(z)=\sum_{N=dist(i,j)}^{\infty} r_{ij}(N) z^{n}  ,  \quad
\overline{r}_{ij}^{(0)}(z)=\sum_{N=dist(i,j)}^{\infty} r_{ij}^{(0)}(N) z^{n} ,
\\
& &\overline{p}_{j}(z)=\sum_{N=2}^{\infty} p_{j}(N) z^{n} ,  \quad
\overline{q}_{j}(z)=\sum_{N=0}^{\infty} q_{j}(N) z^{n} .
\end{eqnarray*}
Multiplying $z^{N}$ and adding in $N$ we have
\begin{equation}
\overline{r}_{ij}(z) =\overline{r}_{ij}^{(0)}(z) + \overline{r}_{ij}(z) \overline{p}_{j}(z)
\label{eqn:y12}
\end{equation}
If $\abs{z} <1$ , $\overline{r}_{ij}(z)$, $\overline{r}_{ij}^{(0)}(z)$ and $\overline{p}_{j}(z)$ converge
absolutely. By definition, 
$$\overline{q}_{j}(z)= \overline{r}_{jj}(z) , \quad  r_{ij}^{(0)} = \overline{r}_{ij}^{(0)}(1) ,$$ 
and the random walk starting from $j$ is transient if and only if $\overline{p}_{j}(1) <1$.
As we assumed that the random walk is uniformly transient, there exists a positive $\epsilon$
such that , for any $j$, $\overline{p}_{j}(1) <1-\epsilon$. As a consequence,
\begin{equation}
\overline{r}_{ij}(1) =\frac{ r_{ij}^{(0)}}{ 1- \overline{p}_{j}(1)} \leq \frac{ r_{ij}^{(0)}}{\epsilon}.
\label{eqn:y13}
\end{equation}
Returning to the Neumann expansion of the resolvent, we have the following relation:
\begin{equation}
r_{N}(ij)  = \left(\delta_{i} ,  (\frac{1}{d}A_{\Gamma})^{N} \delta_{j} \right)_{l^{2}(\Gamma)} .
\label{eqn:y14}
\end{equation}
If $i$ and $j$ are different, we obtain
\begin{equation}
\left(\delta_{i} ,  \frac{1}{- \bigtriangleup_{\Gamma}} \delta_{j} \right)_{l^{2}(\Gamma)} 
 =\sum_{N=0}^{\infty}
 \left(\delta_{i} ,  (\frac{1}{d}A_{\Gamma})^{N}  \frac{1}{d} \delta_{j} \right)_{l^{2}(\Gamma)} 
\leq  \frac{\overline{r}_{ij}(1)}{d(j)}  <  \frac{ r_{ij}^{(0)}}{\epsilon}.
\label{eqn:y15}
\end{equation}
\textbf{End of Proof.}
\\
\\
\begin{lmm}
Set  
\begin{equation}
\underline{\rho} (\beta) =\limsup_{n}  \frac{1}{\abs{\Gamma_{n}}}    
\mathrm{tr}_{l^{2}(\calV)} ( P_{\calV_{n}} \frac{e^{\beta\bigtriangleup_{\Gamma}}}
{1-e^{\beta\bigtriangleup_{\Gamma}}}  P_{\calV_{n}} ) .
\label{eqn:y16}
\end{equation}
Assume that the simple random walk is uniformly transient. Then,  $\underline{\rho} (\beta)$ is finite.
\label{lmm:b4}
\end{lmm}
\textbf{Proof.}
As $-\bigtriangleup_{\Gamma}$ is a positive bounded operator,
the functional calculus implies the following inequality with suitably large $C$.
$$\frac{e^{\beta\bigtriangleup_{\Gamma}}}{1-e^{\beta\bigtriangleup_{\Gamma}}}
\leq C \frac{1}{-\bigtriangleup_{\Gamma}} .$$
By previous lemma, we have
\begin{eqnarray}
& & \:\: \mathrm{tr}_{l^{2}(\calV)} ( P_{\calV_{n}} \frac{e^{\beta\bigtriangleup_{\Gamma}}}
{1-e^{\beta\bigtriangleup_{\Gamma}}}  P_{\calV_{n}} )
\leq C \sum_{i \in \calV_{n}} 
\left( \delta_{i} , \frac{1}{-\bigtriangleup_{\Gamma}}  \delta_{i}\right)_{l^{2}(\calV)}
\nonumber\\
&& \leq \abs{\Gamma_{n}}  \sup_{i \in \calV_{n}}  \overline{q}_{i}(1).
\label{eqn:y17}
\end{eqnarray}
This implies the finiteness of $\underline{\rho} (\beta)$.
\textbf{End of Proof.}
\\
\\
\begin{lmm}
Assume that the simple random walk is uniformly transient. 
Then,
\begin{equation}
\overline{\rho}(\beta) \leq \underline{\rho} (\beta) < \infty .
\label{eqn:y171}
\end{equation}
In particular,$\overline{\rho} (\beta)$ is finite.
\label{lmm:b5}
\end{lmm}
\textbf{Proof.}
Assuming
$\underline{\rho}(\beta) < \overline{\rho} (\beta)$, we show contradiction.
Take $\rho$ satisfying
$\underline{\rho}(\beta) <\rho < \overline{\rho} (\beta)$.
There exists a sequence of subgraphs $\Gamma_{n(k)}$ and $z_{k}$ such that
the following is valid:
\begin{eqnarray} 
&& \rho = tr_{l^{2}(\calV_{n(k)})} (  \frac{ z_{k} e^{\beta\bigtriangleup_{\Gamma_{n(k)}}}}
{1- z_{k} e^{\beta\bigtriangleup_{\Gamma_{n(k)}}}} ) ,
 \nonumber\\
&& \lim_{k} z_{k} = z_{\infty} < 1 .
\label{eqn:y18}
\end{eqnarray}
However, the inequalities (\ref{eqn:y18}) and (\ref{eqn:y7}) imply
that $\rho \leq \underline{\rho}(\beta)$.
\textbf{End of Proof.}
\\
\\
To prove Theorem \ref{Th:main} (ii), we introduce a new notation.
For any positive integer $m$ we define an augmented boundary:
$$\partial_{m} \Gamma_{n} = \left\{  i \in \calV_{n} \: \vert dist ( i , \Gamma_{n}^{c}) \leq m \right\}$$
where $\Gamma_{n}^{c}$ is the complement of $\Gamma_{n}$ in $\Gamma$.
Then, obviously,
$$\abs{\partial_{m} \Gamma_{n} } \leq \overline{d}^{m} \abs{\partial \Gamma_{n}} .$$
As a result,  
$$\lim_{n \to \infty}  \frac{\abs{\partial_{m} \Gamma_{n} }}{\abs{ \Gamma_{n} }} = 0.$$
\begin{lmm}
Suppose that the simple random walk is uniformly recurrent and
fix large $K$ and $n_{0}$ such that the inequality (\ref{eqn:x17b}) is vaild.
Take larger $n_{1}$ satisfying 
$$n_{0}\leq n_{1}  , \quad \frac{\abs{\partial_{n_{o}} \Gamma_{n} }}{\abs{ \Gamma_{n} }} <1/2$$
for $n \geq n_{1}$. 
There exists a constant C independent of  $K$ such that
\begin{equation}
\overline{\rho}(\beta) \geq C K .
\label{eqn:y19}
\end{equation} 
\label{lmm:b6}
\end{lmm}
\textbf{Proof.}
First by use of Trotter Kato formula, we show
\begin{equation}
(\delta_{i} , e^{\beta\bigtriangleup_{\Gamma_{n}} } \delta_{i})_{l^{2}(\calV_{n})}
\geq  
(\delta_{i} , e^{\beta P_{\calV_{n}} \bigtriangleup_{\Gamma} P_{\calV_{n}} }  \delta_{i})_{l^{2}(\calV_{n})}.
 \label{eqn:y20}
 \end{equation}
 As $d_{n}\leq P_{\calV_{n}} d P_{\calV_{n}}$ and 
$A_{\Gamma_{n}} =  P_{\calV_{n}} A_{\Gamma} P_{\calV_{n}}$,
 \begin{eqnarray*}
& &(\delta_{i} ,  e^{\beta\bigtriangleup_{\Gamma_{n}}} \delta_{i})_{l^{2}(\calV_{n})} 
=\lim_{N\to \infty} (\delta_{i} , ( e^{-\frac{\beta d_{n}}{N}} e^{\frac{\beta A_{\Gamma_{n}}}{N}})^{N}
  \delta_{i})_{l^{2}(\calV_{n})}
 \\
 \geq & &
 \lim_{N\to \infty} (\delta_{i} , ( e^{-\frac{\beta P_{\calV_{n}} d P_{\calV_{n}}}{N}} 
 e^{\frac{\beta  P_{\calV_{n}} A_{\Gamma}P_{\calV_{n}} }{N} })^{N}  \delta_{i})_{l^{2}(\calV_{n})} 
 \\
 =&&
 (\delta_{i} , e^{\beta P_{\calV_{n}}\bigtriangleup_{\Gamma} P_{\calV_{n}} }  \delta_{i})_{l^{2}(\calV_{n})} .
\end{eqnarray*}
Thus we obtained (\ref{eqn:y20}) .
\par
Next take $C_{1}$ such that $ e^{\beta x} -1 \leq  C_{1} x $ for $x$ satisfying $0 < x < \overline{d}$.
Note that $e^{x} - z = e^{x}-1 +(1-z)$. By (\ref{eqn:y20}) we have
\begin{eqnarray}
& &\frac{1}{\abs{\Gamma_{n}}} tr_{l^{2}(\calV_{n})} 
( \frac{ z e^{\beta\bigtriangleup_{\Gamma_{n}}}}{1- z e^{\beta\bigtriangleup_{\Gamma_{n}} }} )
\geq C_{1} \frac{1}{\abs{\Gamma_{n}}} 
tr_{l^{2}(\calV_{n})} (\frac{ z}{- \bigtriangleup_{\Gamma_{n}} +1-z})
\nonumber\\
& &\geq C_{1} \frac{1}{\abs{\Gamma_{n}}} 
tr_{l^{2}(\calV_{n})} (\frac{ z}{-  P_{\calV_{n}}\bigtriangleup_{\Gamma}  P_{\calV_{n}}+1-z})
\nonumber\\
& &\geq C_{1} z  \frac{1}{\abs{\Gamma_{n}}} 
\left\{ \sum_{i \in \Gamma_{n}\cap \partial_{n_{1}}\Gamma_{n}^{c}}
(\delta_{i} ,  \frac{1}{ P_{\calV_{n}}(- \bigtriangleup_{\Gamma}) P_{\calV_{n}}+1-z}
\delta_{i})_{l^{2}(\calV_{n})}  \right\} .
\label{eqn:y21}
\end{eqnarray}
As before we use the Neumann expansion of the resolvent:
\begin{eqnarray}
&&(\delta_{i} ,  \frac{1}{ P_{\calV_{n}}(- \bigtriangleup_{\Gamma})P_{\calV_{n}}+1-z}
\delta_{i})_{l^{2}(\calV_{n})}
\nonumber\\
=&& \sum_{L=0}^{\infty}
(\delta_{i} ,  \frac{1}{d+1-z} (  P_{\calV_{n}}A_{\Gamma}  P_{\calV_{n}}  \frac{1}{d+1-z})^{L}
\delta_{i})_{l^{2}(\calV_{n})}
\nonumber\\
\geq && \sum_{L=0}^{n_{1}}
(\delta_{i} ,  \frac{1}{d+1-z} (  P_{\calV_{n}}A_{\Gamma}  P_{\calV_{n}}  \frac{1}{d+1-z})^{L}
\delta_{i})_{l^{2}(\calV_{n})}
\label{eqn:y22}
\end{eqnarray}
as each term in the above Neumann expansion is non-negative.
\\
As far as $i$ belongs to $\Gamma_{n}\cap \partial_{n_{1}}\Gamma_{n}^{c}$ and $L \leq n_{1}$
\begin{eqnarray}
&&(\delta_{i} ,  \frac{1}{d+1-z} (  P_{\Gamma_{n}}A_{\Gamma}  P_{\Gamma_{n}}  \frac{1}{d+1-z})^{L}
\delta_{i})_{l^{2}(\calV_{n})}
\nonumber\\
=&&(\delta_{i} ,  \frac{1}{d+1-z} ( A_{\Gamma} \frac{1}{d+1-z})^{L}
\delta_{i})_{l^{2}(\calV_{n})} .
\label{eqn:y23}
\end{eqnarray}
This is because the random walk starting from $i$ cannot reach the augmented boundary
 $\partial_{n_{1}}\Gamma_{n}$ within $L$ steps.
\\
Now we combine these estimates 
\begin{eqnarray}
&&\frac{1}{\abs{\Gamma_{n}}} tr_{l^{2}(\calV_{n})} 
( \frac{ z e^{\beta\bigtriangleup_{\Gamma_{n}}}}{1- z e^{\beta\bigtriangleup_{\Gamma_{n}} }} )
\nonumber\\
\geq&&
\frac{C_{1} z}{2 \abs{\Gamma_{n}\cap \partial_{n_{1}}\Gamma_{n}^{c}}}
\sum_{ i \in \Gamma_{n}\cap \partial_{n_{1}}\Gamma_{n}^{c}}
\sum_{L=0}^{n_{1}}(\delta_{i} ,  \frac{1}{d+1-z} ( A_{\Gamma} \frac{1}{d+1-z})^{L}\delta_{i})_{l^{2}(\calV_{n})}
\nonumber\\
\label{eqn:y24}
\end{eqnarray}
As a consequence
\begin{eqnarray}
&&\overline{\rho}(\beta )
 \geq \frac{C_{1}}{2}
\inf_{ i \in \calV_{n}\cap \partial_{n_{1}}\Gamma_{n}^{c}}
\sum_{L=0}^{n_{1}}(\delta_{i} ,  \frac{1}{d} ( A_{\Gamma} \frac{1}{d})^{L}\delta_{i})_{l^{2}(\calV_{n})}
\nonumber\\
=&&
\frac{C_{1}}{2} \inf_{ i \in \calV_{n}\cap \partial_{n_{1}}\Gamma_{n}^{c}}  q_{i}(n(1))
\geq \frac{C_{1}}{2} K .
\label{eqn:y25}
\end{eqnarray}
\textbf{End of Proof.}
\\
The above lemma completes our proof of Theorem \ref{Th:main}.
\\
\textbf{End of Proof of  Theorem \ref{Th:main}.}

\newpage
\section{Proof of Theorem\ref{Th:Periodic}.}
\setcounter{theorem}{0}
\setcounter{equation}{0}
In this section, we prove Theorem \ref{Th:Periodic}. As remarked before, 
when the one-particle Hamiltonian is a Schr\"odinger operator on Euclidean spaces, 
the question of occurrence of the Bose-Einstein condensation is reduced to 
behavior of the density of states at the bottom of the spectrum. In case of  periodic potentials
W.Kirsch and B.Simon proved that the asymptotic behavior of the density states
is same as that for the free Schr\"odinger operator. (See  \cite{KirschSimon} .) The same problem
is considered in presence of magnetic field by  P.Briet,  H.D.Cornean, and V.A. Zagrebnov
in \cite{Zagrebnov1}. The basic idea of our proof for the periodic lattice
is same as   \cite{KirschSimon}, however, there appears some difference between
the periodic lattice case and the Euclidean case, which we explain below.
\bigskip
\noindent
\par
Let us recall that the periodic lattice $\Gamma =\{\calV ,\calE\}$ is obtained 
by the fundamental domain $\Gamma_{0} =\{\calV_{0} ,\calE_{0} \}$ and its shift.
The shift is denoted by $\tau_{k}$ as before.
The choice of fundamental domain is not unique and we fix one fundamental
domain $\Gamma_{0} =\{\calV_{0} ,\calE_{0} \}$ satisfying two conditions:
\\
(i) it is a connected sub-graph of $\Gamma$. 
\\
(ii) the conditions $\tau_{k}(i) =j$, and $i,j \in \calV_{0}$ imply $i=j$ and $k=0$.  
\\
Furthermore, without loss of generality, we may assume
\\
(iii) $\Gamma_{0}^{(p)}$ does not possess  a multiple edge.
\\
For our purpose, we may consider a larger block $\Gamma_{n}$ as a fundamental domain and a smaller group  $((2n+1)\bfZ)^{\nu}$ as the shift on our lattice. This is the reason why we can
assume (iii).
\bigskip
\noindent
\par
Now, we identify the vertex set $\calV$ with $\calV_{0} \times \bfZ^{\nu}$
in such a way that the shift on  $\calV_{0} \times \bfZ^{\nu}$ acts via the following formula:
$$\tau_{k}(a,j)=(a,j+k) .$$
$l^{2}(\calV)$ is naturally isomorphic to $l^{2}(\calV_{0}) \otimes l^{2}(\bfZ^{\nu})$, 
We identify this Hilbert space with $l^{2}(\calV_{0}) \otimes L^{2}(T^{\nu})$ by use of
Fourier transformation $F$ from $l^{2}(\bfZ^{\nu})$ to $L^{2}(T^{\nu})$ where
the torus $T^{d}$ is identified with $[-\pi ,\pi ]^{\nu}$. Our convention for Fourier transform
$F$  and the inner product of $L^{2}(T^{\nu})$ are 
$$ ( f, g )_{L^{2}(T^{\nu})} = \frac{1}{(2\pi)^{\nu}}\int_{T^{\nu}}  \overline{f}(p ) g(p ) dp
, \quad
F \delta_{k} = e^{i k\cdot p} \quad \quad k \in \bfZ^{\nu}, p \in  [-\pi ,\pi ]^{\nu}.$$
Then any operator $B$ commuting with shift is called {\em translationally invariant}.
It is unitarily equivalent to a matrix valued multiplication operator $\widetilde{B}(p)$:
$$F B F^{-1}f(p) = \widetilde{B}(p) f(p) \quad \quad , 
f(p) \in l^{2}(\calV_{0}) \otimes L^{2}(T^{\nu}) .$$
Thus, we have direct integral representation of any translationally invariant operator
$B$ : 
$$ B = \int^{\oplus}_{T^{\nu}}  \widetilde{B}(p) dp  , \quad    
l^{2}(\calV ) =  \int^{\oplus}_{T^{\nu}}  {\mathfrak H}_{p} dp$$
where $ {\mathfrak H}_{p}$ is a $\abs{\Gamma_0}$ dimensional Hilbert space
of wave functions satisfying the following twisted boundary condition:
\begin{equation}
{\mathfrak H}_{p} = \{ f(a,j) \in l^{\infty}(\calV) \vert  f(a,j+k) = e^{i k\cdot p} f(a,j) \} .
\label{eqn:z002}
\end{equation} 
The above observation is valid for finite periodic graphs  $\Gamma_{n}^{(p)}$
introduce in Section 1 as well. We have only to replace the $\bfZ^{\nu}$ with the finite cyclic group
 $\bfZ_{n}^{\nu}$ ($\bfZ_{n} = \bfZ /n\bfZ$).
Following the custom of physicists, we call the variable $p$ {\it quasi-momentum}.
\noindent
\bigskip
\par
$\widetilde{A}_{\Gamma}(p)$ is a matrix with entires indexed by
vertices of the fundamental domain $\Gamma_{0}$.
The matrix  elements  $[\widetilde{A}_{\Gamma}]_{ij}(p)$
of $\widetilde{A}_{\Gamma}(p)$ ( for a fixed quasi-momentum $p$) 
are described as follows.
\par
Set $e_{i}(k,j) = \delta_{i,j}$ where $k \in \bfZ^{\nu}$ and $i, j \in \Gamma_0$.  
We regard $e_{i}$ a periodic function on $\calV$.   Obviously $e_{i}$ is in
 $l^{\infty}(\calV )$ and $\{e_{i} (i \in \calV_0 ) \}$ is a basis of
 the set of periodic functions on $\calV$.
When the quasi-momentum $p$ is zero, $[ \widetilde{A}_{\Gamma}]_{ij}(0)$ is determined by
\begin{equation}
A_{\Gamma} e_{j} =\sum_{i \in \calV_0} [\widetilde{A}_{\Gamma}]_{ij}(0) e_{i} .
\label{eqn:z001}
\end{equation}
As the degree of graph is bounded, the adjacency matrix $A_{\Gamma}$ is a bounded operator on 
$l^{\infty}(\calV )$ and (\ref{eqn:z001}) should be understood as an identity of  $l^{\infty}(\calV )$.  
\par
To consider the case  for non vanishing quasi-momentum $p$, we set
$$e_{i}^{(p)}(k,j) =  e^{i k\cdot p} \delta_{ij} .$$
$\{ e_{i}^{(p)}(k,j) \}$ is a basis of the space ${\mathfrak H}_p$ of functions with quasi-momentum $p$.
Then, the matrix elements of $  \widetilde{A}_{\Gamma}(p)_{ij}$ is determined by
\begin{equation}
A_{\Gamma} e_{j}^{(p)} =\sum_{i \in \calV_0} [\widetilde{A}_{\Gamma}]_{ij}(p) e_{i}^{(p)} .
\label{eqn:z003}
\end{equation}
By construction and translational invariance, we obtain the following formulae.
\begin{lmm}
(i) If $i$ and $j$ are connected by an edge inside the fundamental domain $\Gamma_0$,
\begin{equation} 
[\widetilde{A}_{\Gamma}]_{ij}(p) =[\widetilde{A}_{\Gamma}]_{ij}(0) . 
\label{eqn:z004}
\end{equation}
(ii) If $i$ and $j$ are connected by an edge bridging adjacent blocks of $\Gamma$,
\begin{equation}
[\widetilde{A}_{\Gamma}]_{ij}(p)= e^{i \theta_{ij}(p)} [\widetilde{A}_{\Gamma}]_{ij}(0).
\label{eqn:z005}
\end{equation}
 Here $\theta_{ij}(p)$ is real and  $\theta_{ij}(p) = -\theta_{ji}(p)$. It is a linear combination
 of the compnent $p(k)$ of the quasi-momentum $p=(p(1), p(2),\cdots p(\nu))$ described as follows:
If the edge $(i,j)$ of  $\calE_{0}^{(p)}$ corresponds to an edge $(i,j)$ of $\Gamma$ such that
\begin{eqnarray*}
&&i = ( (i_{1}, i_{2},\cdots ,i_{\nu}) , a) \:\:  , \:\:
j = ( (j_{1}, \cdots ,j_{\nu}) , b) \in \calV = {\bfZ}^{\nu} \times \calV_{0} ,
\nonumber\\
&& 1 \leq \sum_{n=1}^{\nu} ( j_{n} - i_{n})  \leq \nu , \quad i_{n}\leq j_{n} \:\: ( n=1,2,\cdots ,\nu ) ,
\end{eqnarray*}
\begin{equation}
\theta_{ij}(p) = \sum_{n=1}^{\nu}  p_{n} ( j_{n} -i_{n}) . 
\label{eqn:z121}
\end{equation}
(iii) If $(i,j)$ is not an edge of $\Gamma$,
\begin{equation}
[\widetilde{A}_{\Gamma}]_{ij}(p) = 0.
\label{eqn:z006}
\end{equation}
\label{lmm:c0}
\end{lmm}
For later convenience, we set
$\theta_{ij}(p) = 0$ when  $i$ and $j$ are connected by an edge inside the fundamental domain $\Gamma_0$ and we can write
\begin{equation}
[\widetilde{A}_{\Gamma}]_{ij}(p)= e^{i \theta_{ij}(p)} [\widetilde{A}_{\Gamma}]_{ij}(0)
\label{eqn:z007}
\end{equation}
for any $i$ and $j$. The factor $e^{i \theta_{ij}(p)} $ corresponds to an external magnetic field
in physics and to a curvature in context of the discrete geometry of graphs.
\noindent
\bigskip
\par
Let us return to the Bose-Einstein condensation.  
Let $l$ be a positive integer and we consider $\Gamma_l$ of (\ref{eqn:x300})
and  the finite graph $\Gamma_l^{(p)}$ obtained by  the periodic boundary condition. 
Consider the adjacency matrices $A_{\Gamma}$ and $A_{\Gamma_{l}^{p}}$.
Let $v$ be the periodic potential of Theorem \ref{Th:Periodic}.
$v$ is a periodic potential for the finite periodic system on $\Gamma_l^{(p)}$ as well.
Set
\begin{equation}
h_{l} = E^{(l)} -A_{\Gamma_l^{(p)}} + v ,\quad h = E -A_{\Gamma} + v,
\label{eqn:z1}
\end{equation}
where  $E^{(l)}$ (resp. $E$) is the supremum of the spectrum of
$A_{\Gamma_l^{(p)}} - v$ (resp. $A_{\Gamma} - v$).

Now we consider the particle density. Set
$$\rho_{l}(z) = \frac{1}{\abs{\Gamma_{l}^{(p)}}}
\mathrm{tr}_{l^{2}(\Gamma_{l}^{(p)})}
\left( \frac{ ze^{-\beta h_{l}} }{ 1 - ze^{-\beta h_{l}} }   \right) .$$
 Due to translational invariance,
\begin{equation}
\rho (z) = \lim_{l\to \infty} \rho_{l}(z)
= \frac{1}{(2\pi)^{d}} \int_{ [-\pi ,\pi ]^{d}}
\frac{1}{\abs{\Gamma_{0}} } \mathrm{tr}_{l^{2}(\calV_{0})}
\left( \frac{ ze^{-\beta \widetilde{h}(p)} }{ 1 - ze^{-\beta \widetilde{h}(p)} }   \right)
dp .
\label{eqn:z2}
\end{equation}
As there exist constants $C_{\beta}^{(1)}$ and $C_{\beta}^{(2)}$such that
$$C_{\beta}^{(1)}   \frac{1}{x + 1-z} \leq \frac{ ze^{-\beta x} }{ 1 - ze^{-\beta x} } 
\leq C_{\beta}^{(2)} \frac{1}{x}$$
for positive $x$ we have the following lemma.
\begin{lmm}
Let $E(p)$ be the largest eigenvalue of the matrix $\widetilde{A}_{\Gamma}(p) -\widetilde{v}(p)$.  
\\
(i) Suppose the following integral is finite.
\begin{equation} 
 \int_{ [-\pi ,\pi ]^{\nu}}  \frac{1}{E-E(p)} dp < \infty .
\label{eqn:z3}
\end{equation}
Then $\rho (1)$ is finite and
for any $z_{n}$ and $l_{n}$ satisfying $\lim_{n} z_{n} =1$ and $\lim_{n} l_{n} = \infty$,
\begin{equation} 
\lim_{n\to \infty} \rho_{l_{n}}(z_{n}) = \rho (1)
\label{eqn:z4}
\end{equation}
(ii) Suppose that the following limit is infinite.
\begin{equation}
\lim_{z \nearrow 1} \int_{ [-\pi ,\pi ]^{\nu}}  \frac{1}{E-E(p) +1-z} dp = \infty .
\label{eqn:z5}
\end{equation}
Then, for any large positive $\rho$, there exists $z$ ( $0<z<1$) such that
$\rho = \rho (\beta )$.
\label{lmm:c1}
\end{lmm}
Note that  for any periodic potential $v$ 
$$\widetilde{v}(p) = v \vert_{\calV_{0}}$$
Thus by abuse of notation we identify $\widetilde{v}(p)$ and  $v$.
$$\widetilde{v}(p) = v$$
\par
(\ref{eqn:z5}) of  the above Lemma implies the absence of Bose-Einstein
condensation while the case (i) of the above Lemma suggests Bose-Einstein
condensation for the mean particle density $\rho$ greater than $\rho (1)$.  
We show that (\ref{eqn:z5}) holds if the dimension $\nu$ of our periodic lattice is one or two
and that (\ref{eqn:z3}) is finite if   if the dimension $\nu$ is greater than or equal to three. 
\begin{lmm}
\begin{equation}
E = E(0) \geq E(p).
\label{eqn:z6}
\end{equation}
\label{lmm:c2}
\end{lmm}
The proof is same as periodic Schr\"{o}dinger operators on $\bf{R}^{\nu}$.
For the detail of proof, see Chapter XIII,16 of \cite{ReedSimon}.

\begin{lmm}
In a neighborhood of $p=0$,  
\begin{equation}
\abs{E - E(p)} \leq C \norm{p}^{2} = C \sum_{k=1}^{\nu} p_{k}^{2}
\label{eqn:z7}
\end{equation}
\label{lmm:c3}
\end{lmm}
\textbf{Proof.} 
We set
$$\mathrm{Re}(\widetilde{A}_{\Gamma}(p))
= \frac{1}{2} \left( \widetilde{A}_{\Gamma}(p) +(\widetilde{A}_{\Gamma}(-p) \right) .$$
When the quasi-momentum $p$ is sufficiently small,
$\mathrm{Re}(\widetilde{A}_{\Gamma}(p))$ and hence, 
$\mathrm{Re}(\widetilde{A}_{\Gamma}(p))-v$ are Perron-Frobenius positive matrices and we have a positive vector as a unique eigenvector for the largest eigenvalue. Then,
\begin{equation}
E(p) \geq  \sup_{\norm{f}=1} (f , ( \mathrm{Re}(\widetilde{A}_{\Gamma}(p)) -v ) f)_{l^{2}(\calV_{0})}
=   \sup_{\norm{f}=1} (\abs{f} , \{ \mathrm{Re}(\widetilde{A}_{\Gamma}(p)) -v \}
\abs{f})_{l^{2}(\calV_{0})}
\label{eqn:z71}
\end{equation}
where $\abs{f}$ is a vector $l^{2}(\Gamma_{0})$ with the component  $ \abs{f_{j}}$ ($j \in \Gamma_{0}$).
\par
The matrix element of $\mathrm{Re}(\widetilde{A}_{\Gamma}(p))$ is the real part of
(\ref{eqn:z004}) and  (\ref{eqn:z005}). 
$$[\mathrm{Re}(\widetilde{A}_{\Gamma}(p))]_{ij} =\cos \theta_{ij}(p) 
[\mathrm{Re}(\widetilde{A}_{\Gamma}(0))]_{ij} .$$
When  the quasi-momentum $p$ is sufficiently small,
we have a small positive constant $C$ such that
$$\cos \theta_{ij}(p) \geq  (1- C \norm{p}^{2}) .$$
Thus
\begin{eqnarray}
&&\sup_{\norm{f}=1} (\abs{f} , \{ \mathrm{Re}(\widetilde{A}_{\Gamma}(p)) -v \}
\abs{f})_{l^{2}(\calV_{0})} 
\nonumber\\
\geq&& 
 (1- C \norm{p}^{2})
\sup_{\norm{f}=1} (\abs{f} , \{ (\widetilde{A}_{\Gamma}(0)) -v \}
\abs{f})_{l^{2}(\calV_{0})} 
\nonumber\\
=&& (1- C \norm{p}^{2}) E(0) .
\label{eqn:z72}
\end{eqnarray}
This inequality implies (\ref{eqn:z7}) .
\textbf{End of Proof.}
\noindent
\bigskip
\par
Lemma \ref{lmm:c3} shows the divergence of the integral (\ref{eqn:z5})  if $\nu$ is one or two.
Next we consider the case $\nu \geq 3$.
To show Proposition \ref{pro:c4} below, we use a graph analogue of the diamagnetic inequality.
\begin{lmm}
For any function $f$ on $\calV_{0}$,
\begin{equation}
\left( f , e^{- \beta (E- \widetilde{A}(p)+v)}f \right)_{\calV_{0}}
\leq
\left( \abs{f} , e^{-\beta (E- \widetilde{A}(0)+v)} \abs{f} \right)_{\calV_{0}} .
\label{eqn:z9}
\end{equation}
\label{lmm:c7}
\end{lmm}
\textbf{Proof.} 
Recall that the abosulute value of matrix elements of  $\widetilde{A}(p)$ is same as
that of  $\widetilde{A}(0)$
By the Trotter-Kato product formula, we have
\begin{eqnarray} 
&&\left( f , e^{- \beta (E- \widetilde{A}(p)+v)} f \right)_{\calV_{0}} =
\lim_{N\to \infty}  \left( f , \left[ e^{- \frac{\beta}{N} (E+v)} (1+\frac{\widetilde{\beta A}(p)}{N}
\right]^{N}   f \right)_{\calV_{0}} 
\nonumber\\
\leq  &&\lim_{N\to \infty}  \left( \abs{f} , 
\left[ e^{- \frac{\beta}{N} (E+v)} (1+\frac{\beta \widetilde{A}(0)}{N}) \right]^{N}  \abs{f} \right)_{\calV_{0}} 
\nonumber\\
=&& \left( \abs{f} , e^{-\beta (E- \widetilde{A}(0)+v)} \abs{f} \right)_{\calV_{0}} .
\label{eqn:z73}
\end{eqnarray}
\textbf{End of Proof.}
\begin{pro}
If   $E = E(p_{0})$ for some $p_{0} \ne 0$,
there exists a diagonal unitary $W$ on $l^{2}(\calV_{0})$ such that
\begin{equation}
W \widetilde{A}(p)W^{*}  = \widetilde{A}(p-p_{0})
\label{eqn:z8}
\end{equation}
in a neighborhood of $p_{0}$
\label{pro:c4}
\end{pro}
\textbf{Proof.}
Suppose that  $E =E(0)= E(p_{0})$  and $f$ is the unit eigenvector for the largest eigenvalue
of $\widetilde{A}(p)-v$:
$$( \widetilde{A}(p)-v ) f =Ef ,\quad \norm{f}=1.$$
By the diamagnetic inequality,
\begin{equation}
1= \left( f , e^{-\beta (E- \widetilde{A}(p_{0})+v)} f \right)_{\calV_{0}}
\leq  \left( \abs{f} , e^{-\beta (E- \widetilde{A}(0)+v)} \abs{f} \right)_{\calV_{0}} \leq 1
\label{eqn:z74}
\end{equation}
Thus by differentiating (\ref{eqn:z74}),
$$( \widetilde{A}(0)-v ) \abs{f} =E \abs{f} .$$
It turns out that $\abs{f}$ is the Perron Frobenius eigenvector of $ \widetilde{A}(0)-v$
and all the components of $\abs{f}$ are positive.
Now we define $\delta_{a}$ ( $a \in \Gamma_{0}$ ) and a diagonal unitary $V$ 
via the following equations:
$$ f_{a} = e^{i \delta_{a}} \abs{f_{a}}  , \quad
W =  \mathrm{diag}(  e^{-i \delta_{a}}) .$$
By definition,$f = W^{*}\abs{f}$ , $ W v W^{*} = v$ and
\begin{equation}
(W \widetilde{A}(p_{0})W^{*} -v) \abs{f} = E \abs{f}  . 
\label{eqn:z75}
\end{equation} 
We claim that  
 \begin{equation}
 W \widetilde{A}(p)W^{*} = \widetilde{A}(p-p_{0}) .
\label{eqn:z76}
\end{equation}
First, due to  (\ref{eqn:z007}),
\begin{equation}
 [W \widetilde{A}(p)W^{*}]_{ij} = e^{i (\theta_{ij}(p) +\delta_{i} -\delta_{j})} [\widetilde{A}(0)]_{ij} .
\label{eqn:z761}
\end{equation}
Taking the real part of (\ref{eqn:z75}), we have
\begin{equation}
(B(p_{0}) -v) \abs{f} = E \abs{f}  ,
\label{eqn:z77}
\end{equation} 
where
$$B(p) \equiv  \mathrm{Re}(W \widetilde{A}(p_{0})W^{*}).$$
The matrix elements of  $B(p)$ is given by
$$[B(p)]_{ij} = \cos (\theta_{ij}(p) +\delta_{i} -\delta_{j}) [\widetilde{A}(0)]_{ij} .$$
Consider the following inner product:
\begin{equation}
\left( \abs{f} ,  (B(p_{0}) -v) \abs{f} \right)_{l^{2}(\calV_{0})} 
= E \left( \abs{f} , \abs{f} \right)_{l^{2}(\calV_{0})}  =E
\label{eqn:z78}
\end{equation} 
As $\abs{f}$ is the Perron Frobenius eigenvector of $\widetilde{A}(p_{0}) -v$,
we have two identities:
\begin{equation}
\sum_{ij}  \left( \cos (\theta_{ij}(p_{0}) +\delta_{i} -\delta_{j}) [\widetilde{A}(0)]_{ij} - \delta_{ij} \right)
 \abs{f}_{i}  \abs{f}_{j} =E ,
\label{eqn:z79}
\end{equation} 
\begin{equation}
\sum_{ij}  \left( [\widetilde{A}(0)]_{ij} - \delta_{ij} \right) \abs{f}_{i}  \abs{f}_{j} =E 
\label{eqn:z710}
\end{equation} 
As these two identities must be valid simultaneously and all the components of
$\abs{f}$ are non-vanishing, 
$$\cos (\theta_{ij}(p_{0}) +\delta_{i} -\delta_{j})=1 .$$
This implies $\sin (\theta_{ij}(p_{0}) +\delta_{i} -\delta_{j})= 0$ and
\begin{equation}
e^{\sqrt{-1} (\theta_{ij}(p) +\delta_{i} -\delta_{j})} = 
e^{\sqrt{-1} ( \theta_{ij}(p-p_{0}) +(\theta_{ij}(p_{0})+\delta_{i} -\delta_{j}))} = 
e^{\sqrt{-1} ( \theta_{ij}(p-p_{0}))} 
\label{eqn:z711}
\end{equation} 
The equations ({eqn:z761}) and (\ref{eqn:z711}) suggest the claim of Proposition\ref{pro:c4}.
\\
\textbf{End of Proof.}  
\noindent
\bigskip
\newline
The following is a corollary of  Proposition\ref{pro:c4}.
\begin{cor}
(i) The number of $p_{0}$ satisfying  $E = E(p_{0})$ is finite.
\\
(ii) The integral (\ref{eqn:z3}) is finite if and only if the following integral
is finite in a neighborhood of $p=0$:
\begin{equation}
  \int   \frac{1}{E-E(p)} dp 
 \label{eqn:z81}
\end{equation}
\label{cor:c5}
\end{cor}
\begin{rmk}
(i) For Schr{\"o}dinger operators with periodic potential on Euclidean spaces,
$E = E(p)$ holds only at the origin $p=0$ See \cite{KirschSimon}.
\par
As $\widetilde{A}(p)-v$ can be interpreted as a Schr{\"o}dinger operators
with a magnetic field on the finite graph $\Gamma_{0}$, degenracy
$E = E(p)$ can happen even if $p \ne 0$. Such examples are presented 
explicitly by  Yusuke Higuchi and  Tomoyuki Shirai in \cite{Shirai}.
\\
(ii) The above Proposition \ref{pro:c4} and Corollary \ref{cor:c5} are valid for any dimension $\nu$.
\label{rmk:c6}
\end{rmk}
Next we compare $E-E(p)$ when $v$ is the degree of the periodic lattice
(i.e. when $h =d-A = -\bigtriangleup_{\Gamma} $) and other $v$.
Now $E^{d}$, $E^{d}(p)$ stand for $E$ and $E(p)$ when $v$ is the degree $d$
and $E^{v}$, $E^{v}(p)$ for those of other periodic potential $v$.
By definition, $E^{d} =0$.
We also denote $h(v) = E -A_{\Gamma} +v$ and 
$h_{l}(v) = E^{(l)} -A_{\Gamma_{l}^{(p)}} +v$ .  
\begin{lmm}
Let $\Omega = \Omega (i)$($i \in \calV$) and $\Omega^{(l)} = \Omega^{(l)} (i)$
be the positive periodic ground states for $h(v)$ and $h_{l}(v)$
satisfying the following normalization condition:
$$ ( \Omega , \Omega )_{l^{2}(\Gamma_{0})} = \sum_{i \in \calV_{0}}
 \abs{ \Omega (i)}^{2} = \abs{\calV_{0}} , \quad
( \Omega^{(l)} , \Omega^{(l)} )_{l^{2}(\calV_{0})}  = \abs{\calV_{0}} $$
Then,
\begin{equation}
 ( f \Omega , h(v) f \Omega )_{l^{2}(\calV)} = \sum_{(i,j) \in \calE} 
 \abs{f(i) - f(j)}^{2}  \Omega (i) \Omega (j)  
\label{eqn:z10}
\end{equation}
for any $f$ in $l^{2}(\calV)$ and
\begin{equation}
 ( f \Omega^{(l)} , h_{l}(v) f \Omega^{(l)} )_{l^{2}(\calV_{l})} = \sum_{(i,j) \in \calE^{(p)}_{l}} 
 \abs{f(i) - f(j)}^{2}  \Omega^{(l)} (i) \Omega^{(l)} (j)  
\label{eqn:z11}
\end{equation}
for any $f$ in $l^{2}(\calV_{0})$.
\\
If $f$ satisfies the twisted boundary condition with quasi-momentum $p$, we have
\begin{equation}
 ( f \Omega , (E-A_{\Gamma}(p)+v) f \Omega )_{l^{2}(\calV_{0})} = \sum_{(i,j) \in \calE_{0}^{(p)}} 
 \abs{f(i) - e^{i\theta_{ij}(p)}f(j)}^{2}  \Omega (i) \Omega (j)  
\label{eqn:z12}
\end{equation}
where  $h^{(0)}(v)$, and $\calE_{0}^{(p)}$ are obtained by $A_{\Gamma_{0}} -v$ 
and $\calE_{0}$ with the periodic boundary condition.
$\theta_{ij}(p)$ is defined in (\ref{eqn:z121}) when  $i$ and $j$ are connected by an edge
not belonging to $\calE_{0}$ and when $i$ and $j$ are connected by an edge
in $\calE_{0}$, we set $\theta_{ij}(p) =0$.
\label{lmm:c8}
\end{lmm}
\textbf{ Proof.}
To derive (\ref{eqn:z12}) ,  divide (\ref{eqn:z11}) by the volume of $\Gamma_{l}$ and
take the limit  of $l$ to the infinity.
Thus we have only to show the identity (\ref{eqn:z10}) 
which can be obtained by direct calculation as follows.
(c.f. \cite{KirschSimon}.)
\par
First we consider the case of $d=v$. In this case $\Omega (i) =1$ and by definition,
\begin{eqnarray}
&& ( f \Omega ,( d-A_{\Gamma}) f \Omega )_{l^{2}(\calV)} 
= \sum_{(i,j) \in \calE} \left( \abs{f(i)}^{2} +  \abs{f(j)}^{2} - \overline{f}(j) f(i) - \overline{f}(i) f(j) \right)
\nonumber\\
=&& \sum_{(i,j) \in \calE}  \abs{f(i) - f(j)}^{2}  .
\label{eqn:z1201}
\end{eqnarray}
Next we set $w= v-d$. Then ,$ - (d-A_{\Gamma} ) \Omega =  (E+w) \Omega $.
Using (\ref{eqn:z1201}), 
\begin{eqnarray}
&& ( f \Omega , (E+ d-A_{\Gamma}+ w) f \Omega )_{l^{2}(\calV)} 
\nonumber\\
= &&( f \Omega , ( d-A_{\Gamma}) f \Omega )_{l^{2}(\calV)}
 - ( f \Omega , f (d-A_{\Gamma}) \Omega )_{l^{2}(\calV)}
\nonumber\\
=&& \sum_{(i,j) \in \calE}  \abs{f(i) \Omega (i) - f(j) \Omega (j)}^{2}  
\nonumber\\
&&-\sum_{(i,j) \in \calE}   \left\{ \abs{f(i)}^{2} \left( \abs{\Omega (i)}^{2}  - \Omega (i) \Omega (j)
\right) +  \abs{f(j)}^{2} \left( \abs{\Omega (j)}^{2}  - \Omega (i) \Omega (j))\right) \right\}
\nonumber\\
= &&\sum_{(i,j) \in \calE} 
\left\{  \abs{f(i)}^{2} +  \abs{f(j)}^{2} - \left(\overline{f}(i) f(j)+  \overline{f}(j) f(i)\right) \right\} 
\Omega (i)\Omega (j) 
\nonumber\\
 = &&\sum_{(i,j) \in \calE}  \abs{f(i) - f(j)}^{2}  \Omega (i) \Omega (j) .  
\label{eqn:z1202}
\end{eqnarray}
\textbf{End of Proof.}
\newline
\textbf{Proof of Theorem\ref{Th:Periodic}.}
By Lemma \ref{lmm:c8} ,  we have
\begin{eqnarray}
&&E^{v}(0)  - E^{v}(p)=  \inf_{f \in {\mathfrak H}_{p}}
\frac{ \sum_{(i,j) \in \calE^{(p)}_{0}}  \abs{f(i) - f(j)}^{2}  \Omega (i) \Omega (j)  }
{\sum_{i \in \calV_{0}}  \abs{f(i)}^{2}  (\Omega (i))^{2}} ,
\nonumber\\
&&E^{d}(0)- E^{d}(p) = -E^{d}(p)=  \inf_{f \in {\mathfrak H}_{p}}
\frac{ \sum_{(i,j) \in \calE^{(p)}_{0}}  \abs{f(i) - f(j)}^{2} }{\sum_{i \in \calV_{0}}  \abs{f(i)}^{2}}.
\label{eqn:z13}
\end{eqnarray}
Set
$$M =  \frac{\sup_{i}  \Omega (i)}{\inf_{i}  \Omega (i)} .$$
Due to  (\ref{eqn:z13}) , we obtain
\begin{equation}
M^{-1}(E^{d}(0) -E^{d}(p)) \leq E^{v}(0)  - E^{v}(p) \leq M(E^{d}(0) -E^{d}(p)).
\label{eqn:z14 }
\end{equation}
This inequality shows that finiteness of the integral (\ref{eqn:z81})
in a neighborhood of $p=0$ for the discrete Laplacian
is equivalent to that for the Schr\"{o}dinger operator $h$ with a periodic potential $v$. 
This completes our proof of Theorem\ref{Th:Periodic}.
\textbf{End of Proof.}
\newpage
\textbf{Acknowledgment.}
The author would like to thank Dr.Tomoyuki Shirai (Kyushu University)
for discussion on spectrum property of graph Laplacians.


\end{document}